\documentclass[twoside,11pt]{article}
\usepackage{cite}
\usepackage{caption}
\usepackage{subcaption}
\usepackage{adjustbox}
\usepackage{tabularx}
\usepackage{adjustbox}
\usepackage{enumitem}
\usepackage{multirow}
\usepackage{graphicx}
\usepackage{booktabs} 
\usepackage{amsmath,amssymb,amsfonts}
\usepackage[table, usenames, dvipsnames]{xcolor}

\usepackage{jmlr2e}

\newcommand{\noop}[1]{} 

\firstpageno{1}

\begin{document}

\onecolumn 

\begin{description}[labelindent=-1cm,leftmargin=1cm,style=multiline]

\item[\textbf{Citation}]{Y. Logan, M. Prabhushankar and G. AlRegib, "DECAL: DEployable Clinical Active Learning," ICML 2022 Workshop on Adaptive Experimental Design and Active Learning in the Real World, July 2022} \\


\item[\textbf{Review}]{Date of publication: 22 July 2022} \\

\item[\textbf{Codes}]{\url{https://github.com/olivesgatech/Patient-Aware-Active-Learning} \\}

\item[\textbf{Bib}] {@ARTICLE\{Logan2022\_ICML\_ReALML,\\ 
author=\{Y. Logan, M. Prabhushankar and G. AlRegib\},\\ 
journal=\{ICML 2022 Workshop on Adaptive Experimental Design and Active Learning in the Real World\},\\ 
title=\{DECAL: DEployable Clinical Active Learning\}, \\ 
year=\{2022\}\\ 
} 


\item[\textbf{Contact}]{\href{mailto:ylogan3@gatech.edu}{ylogan3@gatech.edu}  OR \href{mailto:alregib@gatech.edu}{alregib@gatech.edu}\\ \url{http://ghassanalregib.info/} \\ }
\end{description}

\thispagestyle{empty}
\newpage
\clearpage
\setcounter{page}{1}

\jmlrheading{1}{}{}{}{}{}

\title{DECAL: DEployable Clinical Active Learning}

\author{\name Yash-yee Logan 
        \email ylogan3@gatech.edu \\
       \name Mohit Prabhushankar 
       \email mohit.p@gatech.edu \\
       \name Ghassan AlRegib \email 
       alregib@gatech.edu \\
       \addr OLIVES Lab, Georgia Institute of Technology\\
       Atlanta, GA 30332, USA}


\maketitle

\begin{abstract}%
Conventional machine learning systems that operate on natural images assume the presence of attributes within the images that lead to some decision. However, decisions in medical domain are a resultant of attributes within medical diagnostic scans and electronic medical records (EMR). Hence, active learning techniques that are developed for natural images are insufficient for handling medical data. We focus on reducing this insufficiency  by designing a deployable clinical active learning (DECAL) framework within a bi-modal interface so as to add practicality to the paradigm. Our approach is a \emph{plug-in} method that makes natural image based active learning algorithms generalize better and faster. We find that on two medical datasets on three architectures and five learning strategies, DECAL increases generalization across 20 rounds by approximately 4.81\%. DECAL leads to a 5.59\% and 7.02\% increase in average accuracy as an initialization strategy for optical coherence tomography (OCT) and X-Ray respectively. Our active learning results were achieved using 3000 (5\%) and 2000 (38\%) samples of OCT and X-Ray data respectively.

\end{abstract}

\begin{keywords}
  Active Learning, Clinical Context, Real-World Deployment
\end{keywords}

\section{Introduction and Related Work}
Active learning aims to find the optimal subset of samples from a dataset for a machine learning model to learn a task well (\cite{settles2009active, dasgupta2011two}). It is studied because of its ability to reduce the costly and laborious burden on experts to provide data annotations. Typical setups focus on acquisition functions that measure the informativeness of samples using constructs from ensemble learning (\cite{beluch2018power}), probabilistic uncertainty (\cite{hanneke2014theory, gal2017deep}) and data representation (\cite{sener2017active, geifman2017deep}). These works were originally developed for the natural image domain and although several studies have adapted these and other techniques to medical imagery (\cite{shi2019active, nath2020diminishing, melendez2016combining, otalora2017training, logan2022patient}), they have not been adopted or utilized in real clinical settings. 

One reason for this non-adoption is that conventional active learning does not follow the diagnostic process. This is because of the experimental settings in natural images that aided the development of existing active learning algorithms (\cite{ash2019deep, sener2017active, hsu2015active}). Natural images typically contain homogeneous class attributes that can be extracted from the images themselves. Also, these attributes are usually enough to distinguish between classes. However, in medicine, pathologies manifest themselves in visually diverse formats across multiple patients. For example, the characteristics of an aged healthy person are visually different from a young healthy person. So how do doctors overcome this? They include clinical data from EMR to assist with their arrival at a diagnostic decision (\cite{brush2017expert, brundin2018secondary}). EMR consists of patient ID, demographics, diagnostic imaging and test results that allow a clinician to make a diagnosis. We recommend that active learning frameworks for medical image classification be designed within a bi-modal interface so as to add practicality to the paradigm. With this in mind, we design and evaluate a DECAL framework that integrates EMR data. We show that DECAL aids existing active learning algorithms in finding the best subset for labeling as well as initializing the active learning framework. As such, DECAL is a \emph{plug-in} approach on top of existing active learning based methods.

\section{Assessing Active Learning Framework for Medical Domain}
We conduct a set of controlled experiments to evaluate the effectiveness of a DECAL framework relative to conventional frameworks. 

\subsection{Dataset Descriptions}
We use images and EMR data from the OCT dataset by \cite{kermany2018identifying}. The dataset consists of grayscale, cross-sectional, foveal scans having varying sizes. We use images from 3 retinal diseases: 10488 choroidal neovascularization (CNV), 36345 diabetic macular edema (DME) and 7756 Drusen annotated at the image level. Samples in training and oracle sets are from 1852 unique patients. The test set consists of 250 images from each diseased class from 486 unique patients.

We also use images and EMR data from the X-Ray dataset also by \cite{kermany2018identifying}. The X-rays are grayscale, cross-sectional chest scans from children belonging to a healthy class and 2 types of pneumonia: viral and bacterial annotated at the image level. We use 1349 healthy, 1345 viral and 2538 bacterial samples in the combined training and oracle sets from 2650 unique patients. The test set consists of 234 healthy, 148 viral and 242 bacterial images from 431 unique patients. There is also no overlap in patients or imagery in train or test sets for both datasets. This means the imagery in the train and test sets come from different patient cohorts. EMR data used for our analysis was patient identity from both datasets. 

\begin{figure}[h!]
\small
\centering
\includegraphics[width =0.6\linewidth]{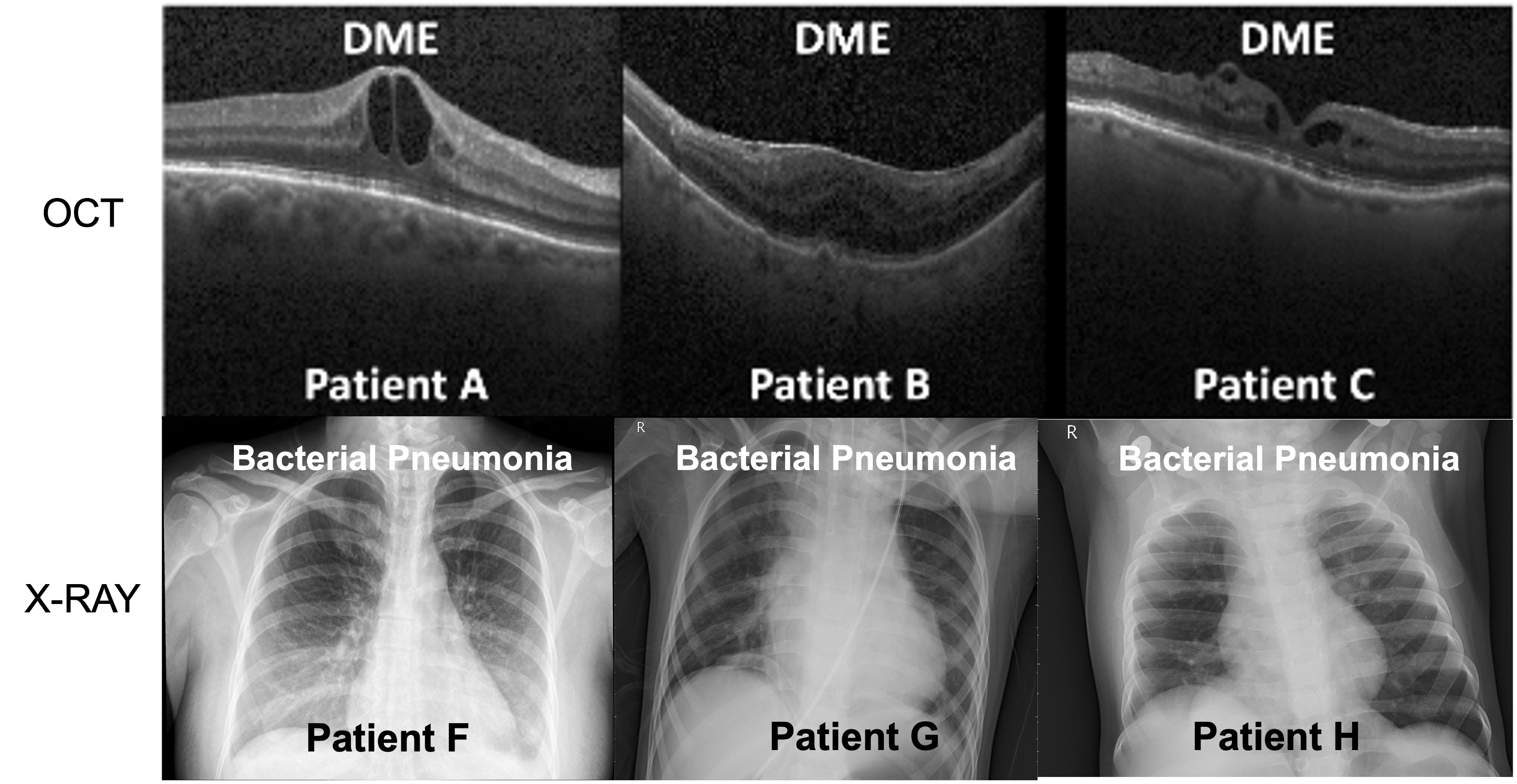}
\caption{Sample imagery of visual characteristics of disease states across patient cohorts.}
\label{fig:diseaseDiversity}
\end{figure}

\vspace{-3mm}
\begin{table}[h!]   
  \begin{center}
    \caption{Information about dataset size and the number of samples added after each training round.}
    \label{tab:implementation details}
    \begin{tabular}{*{13}{c|c|c}}
      \toprule 
      \multirow{2}{*}{\textbf{Details}} & \multicolumn{2}{c}{\textbf{Dataset}} \\
       & OCT & X-Ray \\
      \hline
      \midrule 
      Oracle + Training Set Images  & 54589 & 5232 \\
      Images in Initial Training Set & 128 & 128 \\
      Images Queried per Iteration & 128 & 128 \\
      Unique Patients per Iteration & 128 & 128 \\
      Total Unique Patients & 2338 & 3081 \\
      \bottomrule 
    \end{tabular}
  \end{center}
\end{table}
\subsection{Active Learning with EMR Data}

Figure \ref{fig:diseaseDiversity} shows sample scans from each dataset with patients having the same disease. The visual characteristics across patients are noticeably different. This intra-class diversity is a typical occurrence in medical datasets. Existing active learning paradigms fail to properly account for disease manifestations from whom there is less data. This oversight is very dangerous for safety critical domains like medicine. Thus, we posit that EMR data, in the form of patient identity, be leveraged to account for the intra-class diversity present in medical datasets. We use patient identity as a \emph{plug-in} constraint that can be applied prior to sample selection with any query acquisition function.  The next batch of informative samples will have unique patient identity from the unlabeled pool and be appended to the training set. This process is repeated to determine the minimum number of labeled samples needed to maximize model performance. 

\subsection{Experiments}
\paragraph{Implementation Details}
We assess our active learning framework on Resnet-18, Resnet-50 and Densenet-121 (\cite{he2016deep, huang2017densely}). We do not use pre-trained models in any of our analysis. We use the Adam optimizer with a learning rate of 1.5e-4. Hyper-parameters are tuned based on the OCT dataset and then the same parameters are used for the X-Ray dataset. For each round, the Resnet and Densenet models are trained until 98\% and 94\% accuracy is achieved on the training set respectively. Following each round, the model's weights are reset and randomly initialized. This is repeated with three different random seeds. We aggregate and report average accuracy and standard deviation. All images are resized to $128 \times 128$ and OCT scans are normalized with $\mu=0.1987$ and $\sigma=0.0786$  while X-Rays with $\mu=0.4823$ and $\sigma=0.0379$. Table \ref{tab:implementation details} shows more implementation details for each dataset.

\subsubsection{Initializing Active Learning with EMR Data}
Existing frameworks typically start active learning by randomly selecting a small amount of samples to train the initial model. Subsequently, they apply methods of ranking sample informativeness. By doing this they naively assume that the data distribution is even, which is hardly the case in medical datasets as shown in Figure \ref{fig:diseaseDiversity}. Randomly selecting from an unbalanced distribution is not guaranteed to gather a representative sample of the classes present (\cite{zhu2008active}). Therefore, we recommend the integration of EMR data from the outset to circumvent this. 

To do this, we first compute the distribution of patients throughout the unlabeled pool. Then, we select a fixed number of images from unique patients IDs and pair it with its annotation for the initial training set. The intuition behind this strategy is for the first training samples to have maximally dissimilar images. These samples are then used to start our DECAL paradigm. We present two experimental modalities in the initialization phase depending upon the availability of data.

\paragraph{Large Initial Training Set}
We select 1000 samples at random from the unlabeled pool and train a model for each architecture and dataset for the first round only as our baseline. Then, we perform DECAL initialization by selecting one image from a 1000 unique patients in the unlabeled pool. We then train a model for each architecture and dataset for the first round only and compare it to the baseline by reporting the average accuracy and standard deviation on the test set. The results are presented in Section \ref{subsec:init} Table \ref{tab:OCT initialization}.

\paragraph{Small Initial Training Set}
We select 128 samples with DECAL initialization then start both conventional active learning and DECAL methods and record the earliest round where average accuracy is greater than random chance (33\%). Next we compute the percentage increase/decrease that DECAL achieves relative to the corresponding baseline. The results are presented in Section \ref{subsec:init} Table \ref{tab:ablation}.

\subsubsection{Baseline Sample Acquisition Algorithms}
We apply patient ID as a modular "plug-in" constraint prior to sample selection with each of these baseline algorithms to make our framework clinically deployable. The first baseline is standard random sampling, the next three are margin, least confidence and entropy uncertainty based sampling (\cite{settles2009active}) and the last is an amalgamation of diversity and uncertainty-based sampling approaches known as BADGE (\cite{ash2019deep}). 

\section{Results}
\paragraph{Initializing Active Learning with EMR Data}
\label{subsec:init}
First, we show results to validate the importance of integrating patient ID from the onset. When a large initial train set is used, Table \ref{tab:OCT initialization} shows in yellow that DECAL initialization always leads to higher accuracy regardless of architecture or dataset type. DECAL initialization lead to a 5.59\% and 7.02\% increase in average accuracy for OCT and X-Ray data respectively. However, evaluating the impact of DECAL initialization with a small initial training set mandates a different approach. Since we use non-pre-trained models, training with a small set will unsurprisingly result in over-fitting. This means generalization on the test set will be that of random chance until the training pool becomes large enough. Table \ref{tab:ablation} highlights in yellow the instances when DECAL initialization lead to a percentage increase in average accuracy during early rounds of training. From these results we see DECAL initialization leads to better generalization for at least 5 of the 10 query strategies across all architectures and datasets.




\paragraph{Active Learning with EMR Data}

\begin{figure} [h!]
     \centering
     \begin{subfigure}[b]{0.24\columnwidth}
         \centering
         \includegraphics[width=\columnwidth]{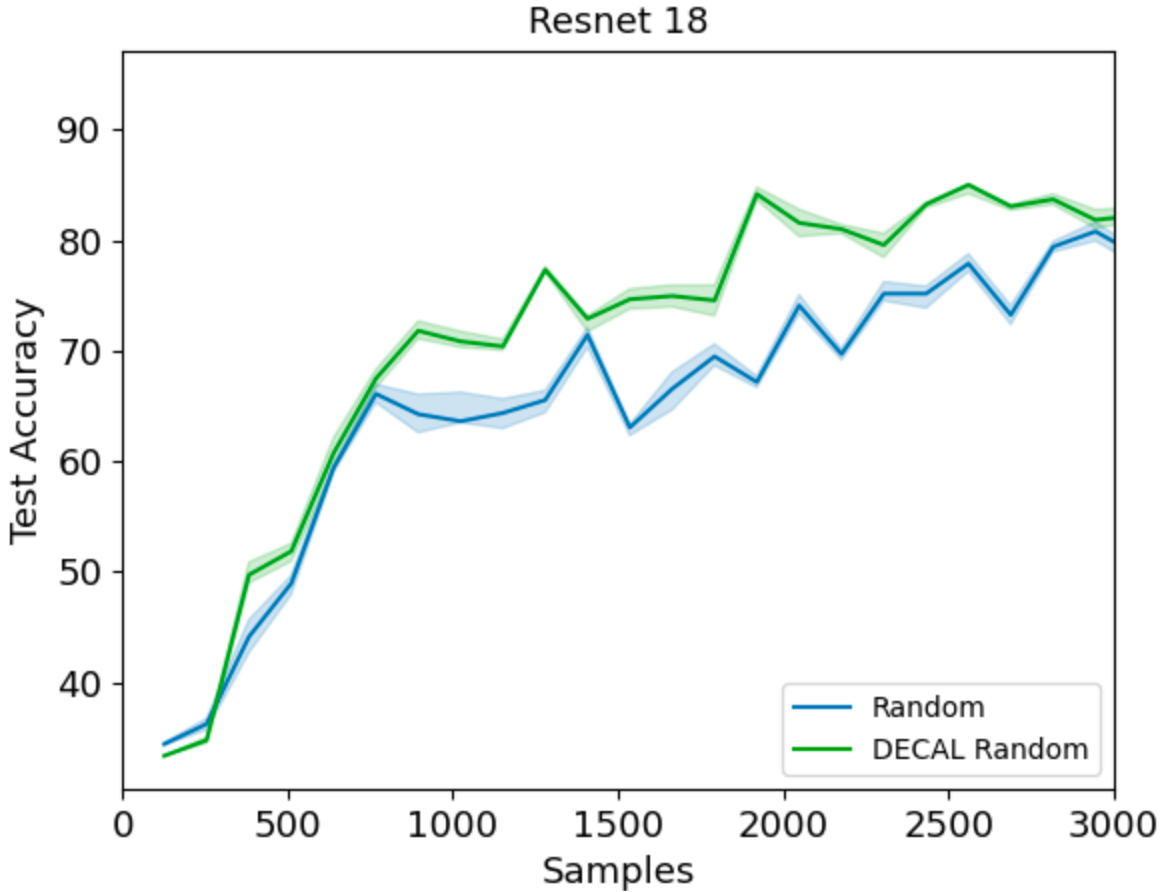}
         \caption{}
         \label{fig: oct rand}
     \end{subfigure}
     \begin{subfigure}[b]{0.24\columnwidth}
         \centering
         \includegraphics[width=\columnwidth]{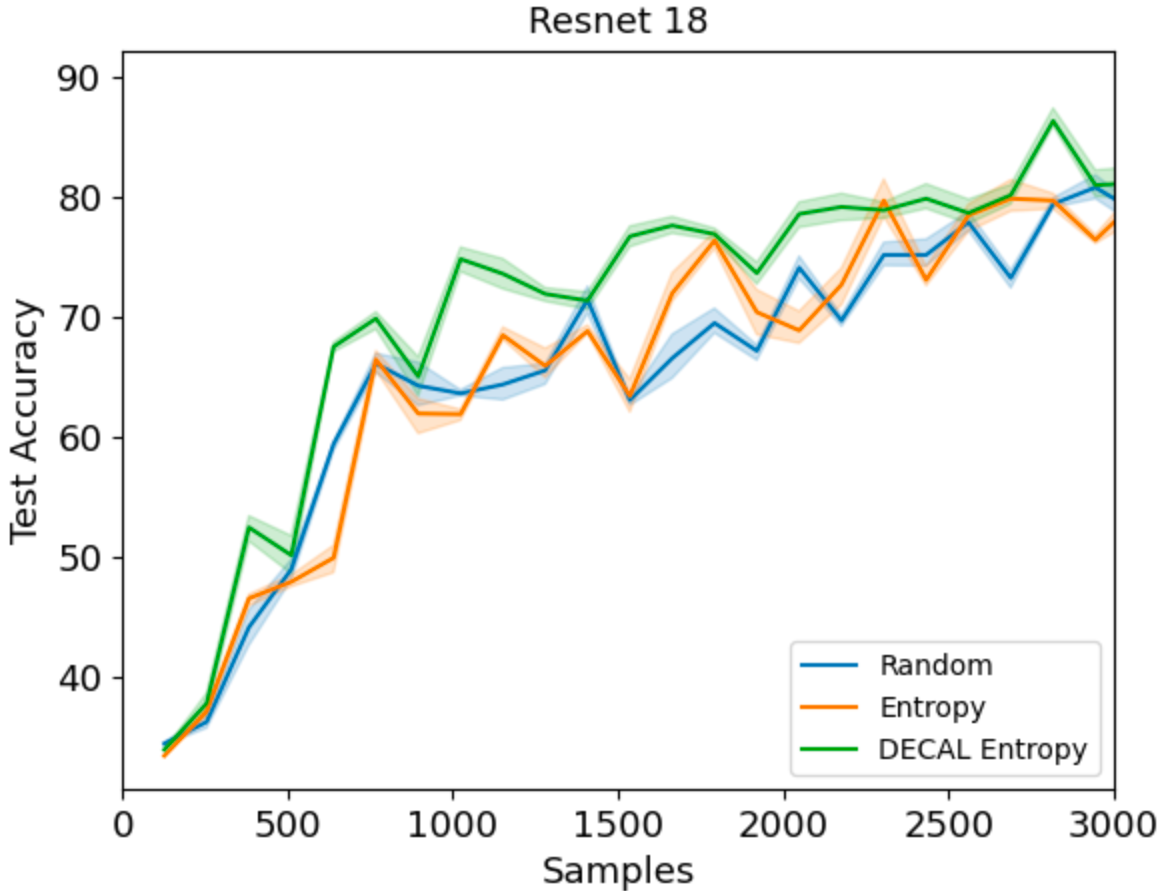}
         \caption{}
         \label{fig:oct entropy}
     \end{subfigure}
     \hfill
     \begin{subfigure}[b]{0.24\columnwidth}
         \centering
         \includegraphics[width=\columnwidth]{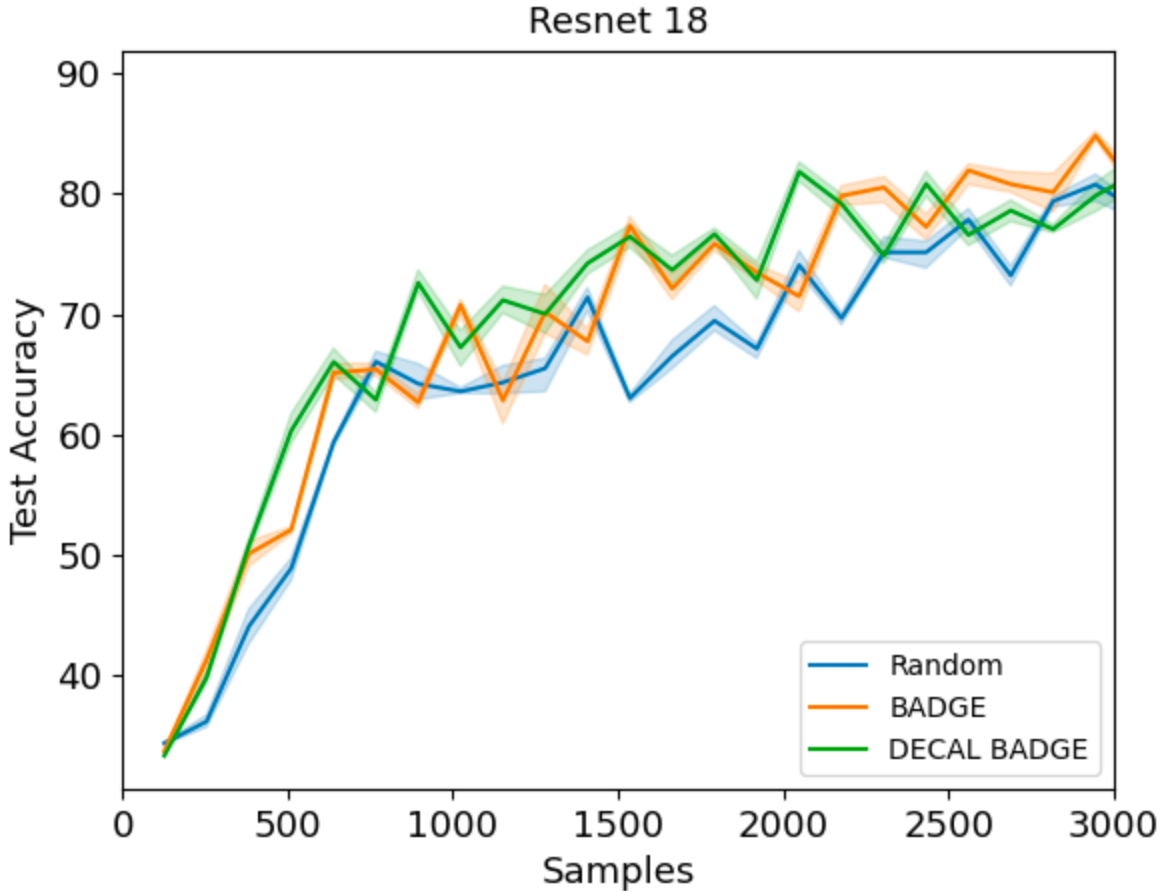}
         \caption{}
         \label{fig:five over x}
     \end{subfigure}
     \hfill
     \begin{subfigure}[b]{0.24\columnwidth}
         \centering
         \includegraphics[width=\columnwidth]{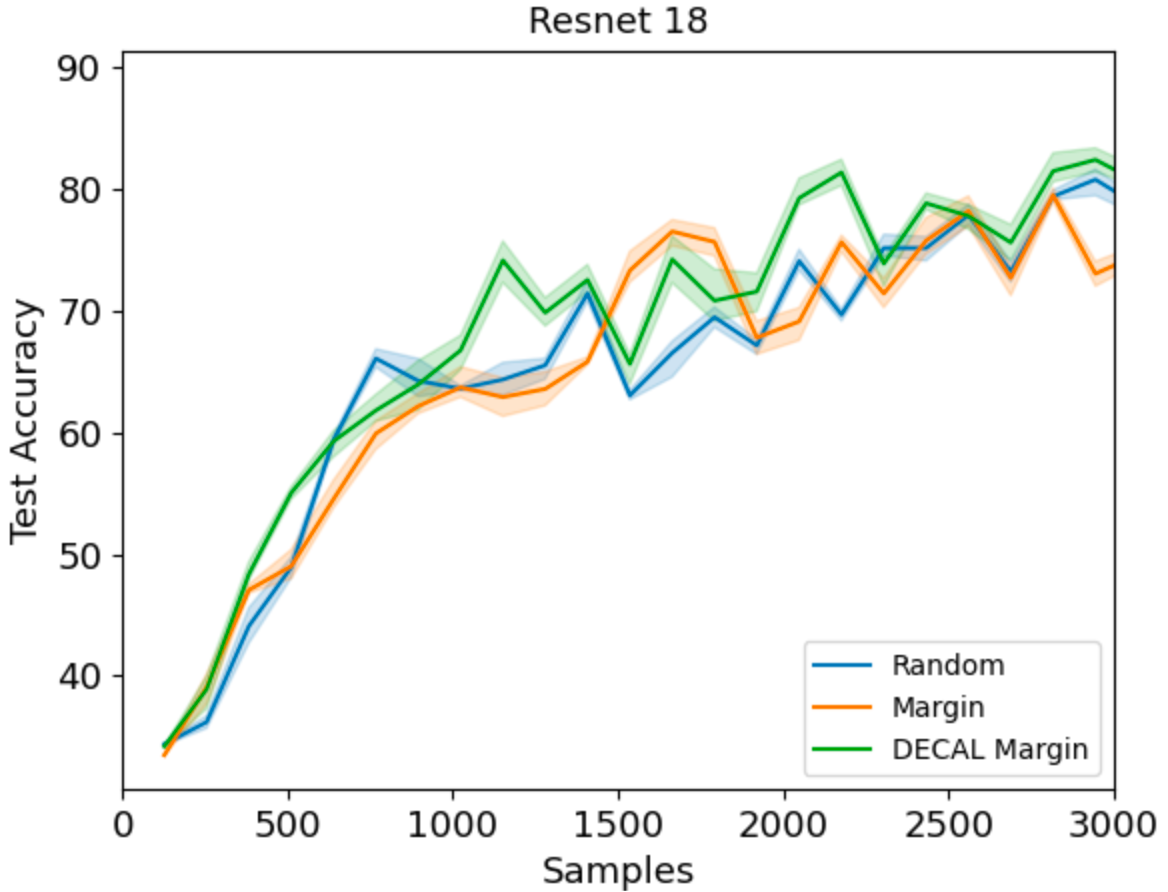}
         \caption{}
         \label{fig:five over x}
     \end{subfigure}
        \caption{Test accuracy vs sample count during training for the OCT on Resnet-18.}
        \label{fig:OCT curves}        
\end{figure}

\begin{figure} [h!]
     \centering
    \begin{subfigure}[b]{0.24\columnwidth}
         \centering
         \includegraphics[width=\columnwidth]{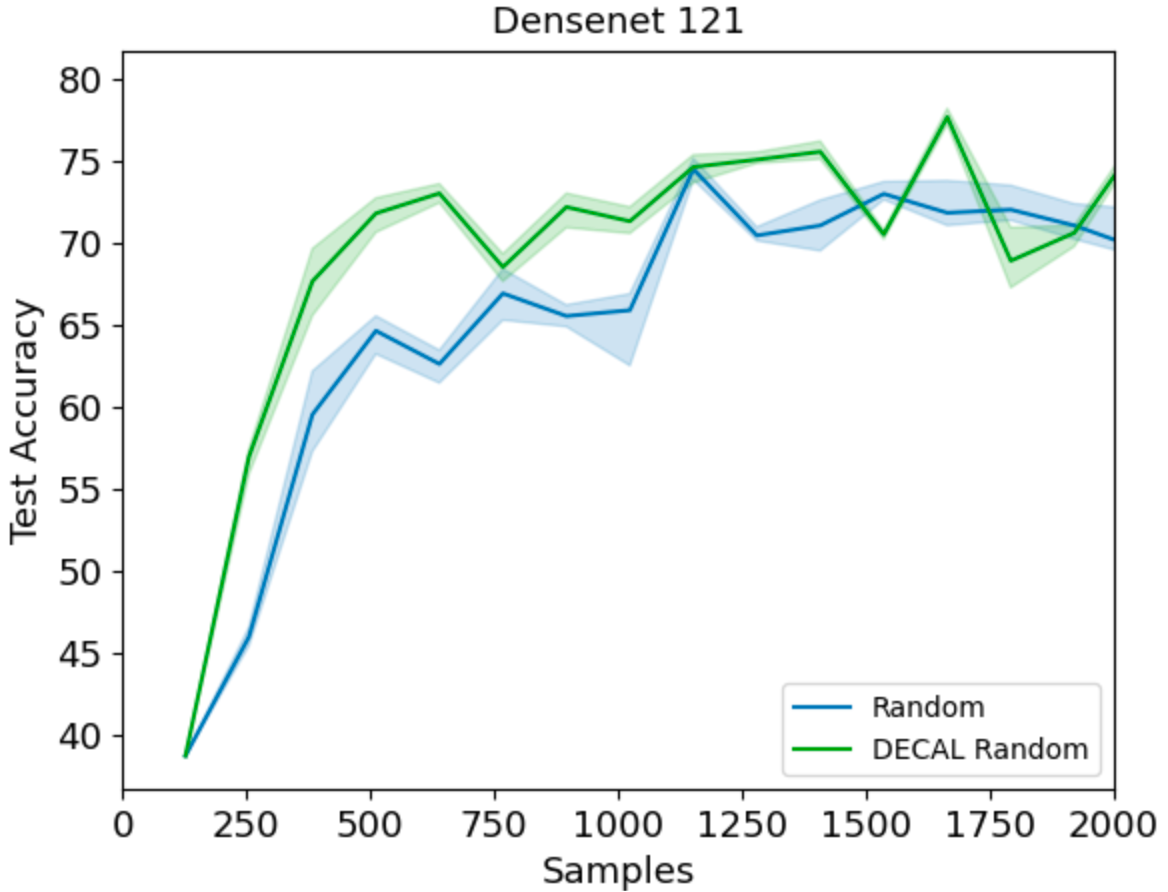}
         \caption{}
         \label{fig:xray rand}
     \end{subfigure}
     \begin{subfigure}[b]{0.24\columnwidth}
         \centering
         \includegraphics[width=\columnwidth]{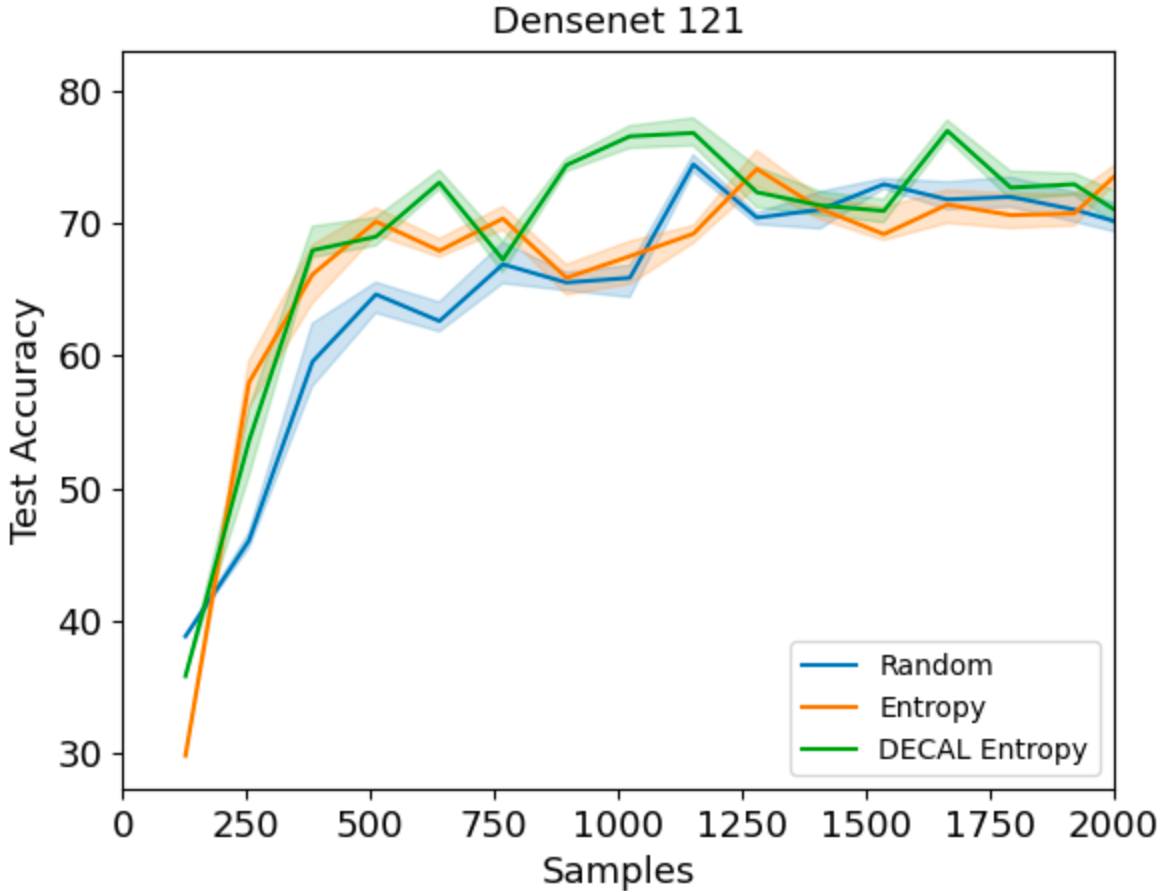}
         \caption{}
         \label{fig:three sin x}
     \end{subfigure}
     \hfill
     \begin{subfigure}[b]{0.24\columnwidth}
         \centering
         \includegraphics[width=\columnwidth]{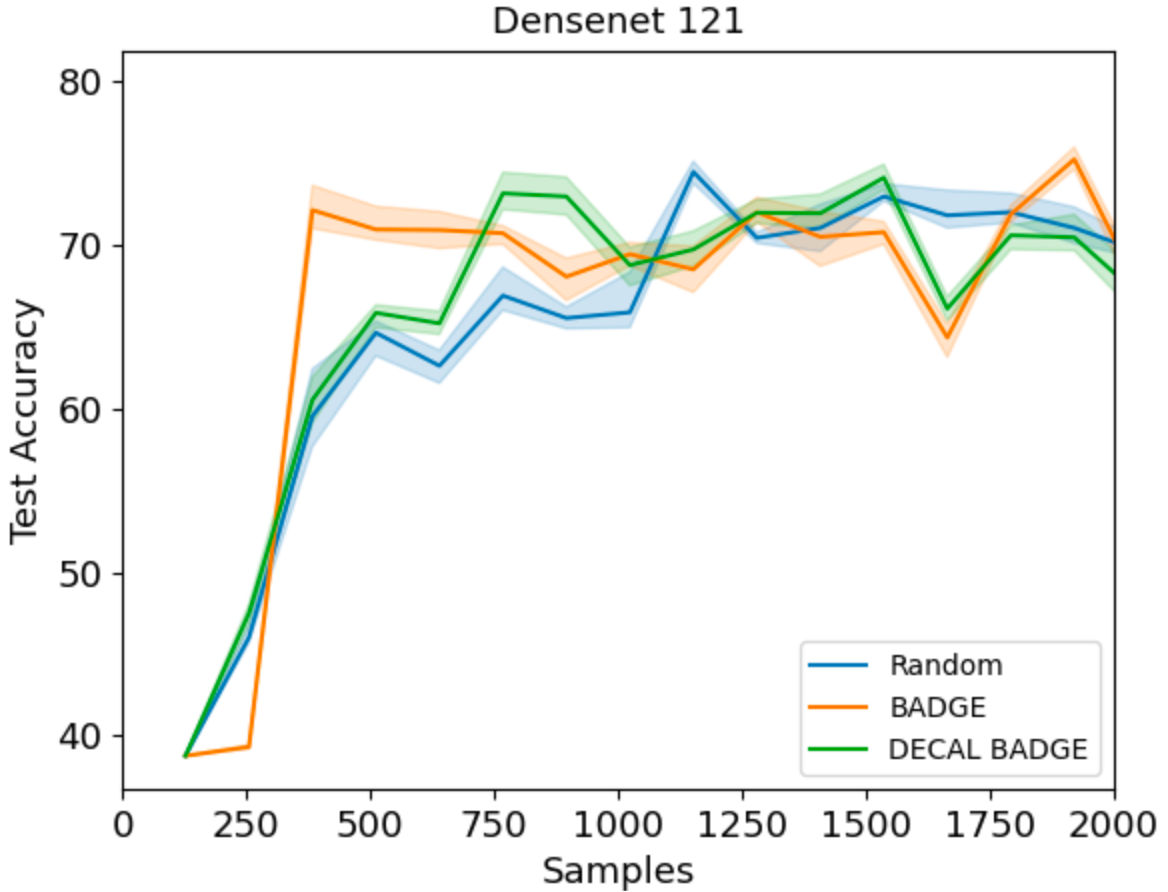}
         \caption{}
         \label{fig: xray badge}
     \end{subfigure}
     \hfill
     \begin{subfigure}[b]{0.24\columnwidth}
         \centering
         \includegraphics[width=\columnwidth]{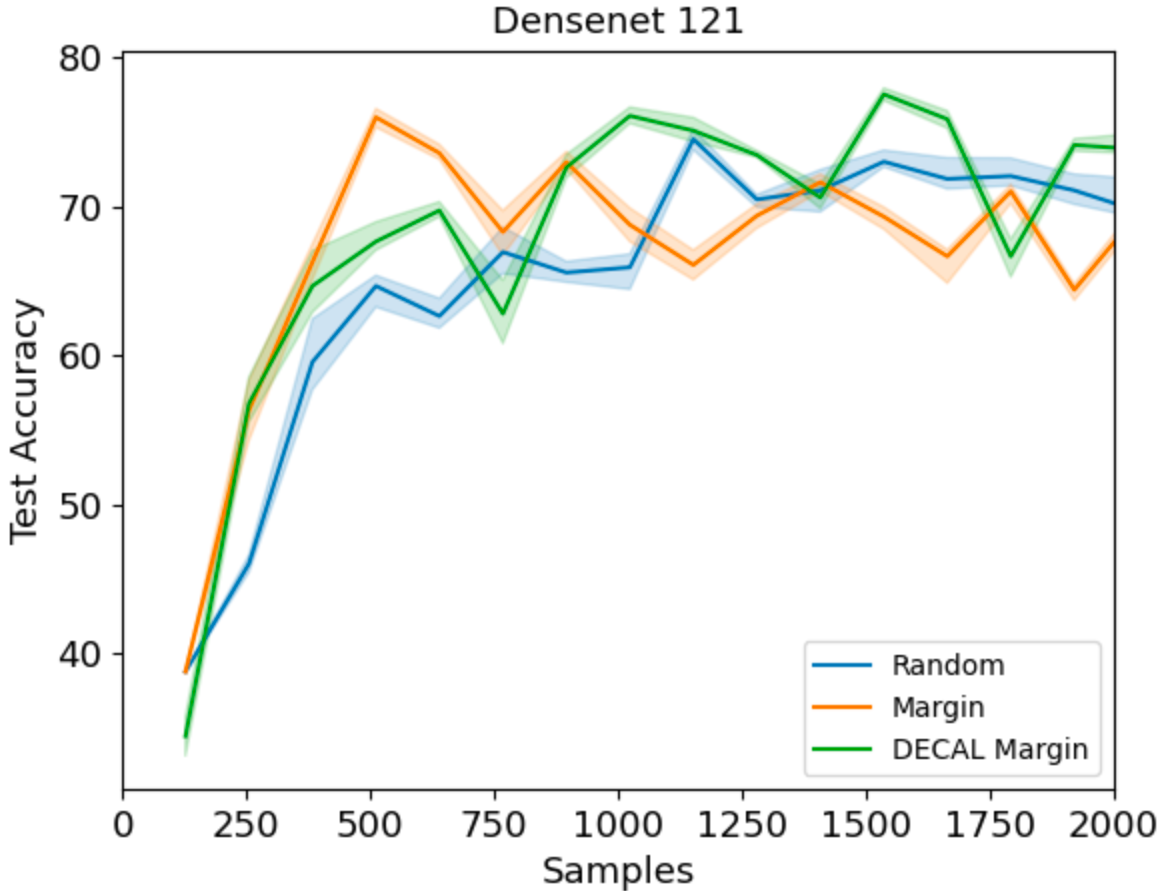}
         \caption{}
         \label{fig:five over x}
     \end{subfigure} 
        \caption{Test accuracy vs sample count during training for the X-Ray on Densenet-121.}
        \label{fig:XRAY curves}
\end{figure}

We use learning curves to evaluate how DECAL can better characterize disease states. Due to space constraints, we show learning curves for only Resnet-18 and Densenet-121 architectures in Figure \ref{fig:OCT curves} and Figure \ref{fig:XRAY curves}. In these plots \texttt{x-axis} corresponds to the number of samples in the train set and \texttt{y-axis} corresponds to the performance accuracy on the test set. Each colored curve is the average of five trials, with standard errors being shown by the shaded regions. We see that DECAL consistently matches or surpasses the baseline algorithms. Furthermore, we see DECAL often having an edge over baseline algorithms from the early rounds of training onward like in Figures \ref{fig:oct entropy}, \ref{fig:xray rand}. This is an indicator that DECAL methods not only generalize better but also generalize faster than the baselines. 
\vspace{-3mm}
\begin{table*}[h!]  
  \begin{center}
    \caption{DECAL vs random initialization using large train set.}
    \label{tab:OCT initialization}
    \begin{adjustbox}{width=0.55\textwidth}
    \begin{tabular}{*{13}{c|c|c|c}}
      \toprule 
      \multicolumn{4}{c}{\textbf{OCT Dataset}}  \\
      \hline
      \midrule
      Initialization & Resnet-18 & Resnet-50 & Densenet-121\\
      \hline
      \midrule 
     Random  & 61.38 $\pm$ 4.53 &	64.4 $\pm$ 6.75 &	76.93 $\pm$ 4.64\\
     DECAL & \cellcolor{yellow} \textbf{64.53 $\pm$ 9.83} & \cellcolor{yellow}	\textbf{69.28 $\pm$ 3.41} &	\cellcolor{yellow} \textbf{80.16 $\pm$ 5.34} \\
     \midrule
     \multicolumn{4}{c}{\textbf{X-Ray Dataset}}  \\
     \hline
     \midrule
     Initialization & Resnet-18 & Resnet-50 & Densenet-121\\
     \hline
     \midrule
     Random  & 52.24 $\pm$ 9.47 & 66.95 $\pm$ 3.98 & 70.12 $\pm$ 7.44\\
     DECAL & \cellcolor{yellow} \textbf{58.2 $\pm$ 9.42} &	\cellcolor{yellow} \textbf{71.08 $\pm$ 4.17} & \cellcolor{yellow} \textbf{72.82 $\pm$ 7.44} \\
      \bottomrule 
    \end{tabular}
    \end{adjustbox}
  \end{center}
\end{table*}
\vspace{-3mm}

\begin{table}[h!]
  \begin{center}
    \caption{DECAL vs random initialization during early rounds of training. +/- numbers show the percentage increase/decrease in average accuracy relative to random.}
    \label{tab:ablation}
    \begin{adjustbox}{width=\textwidth}
    
    \begin{tabular}{*{1}{c|c|c|c|c|c|c|c}}
      \toprule 
       \multirow{3}{*}{} & Dataset & \multicolumn{3}{c}{OCT} \vline & \multicolumn{3}{c}{X-Ray} \\
       & Round & 4 & 2 & 1 & 2 & 2 & 1\\
       & Query & Resnet-18 & Resnet-50 & Densenet-121 & Resnet-18 & Resnet-50 & Densenet-121 \\
      \hline
      \midrule 
        \parbox[t]{2mm}{\multirow{10}{*}{\rotatebox[origin=c]{90}{Random Initialization}}} & Random & 54.37 & 46.18 & 42.66 & 63.91 & 69.64 &	46.37\\
      & Entropy & 53.62	& 46.88 & 49.97 & 55.88 &	51.18 &	46.18\\
      & BADGE & 60.53 &	48.69 &	35.3 & 57.50 &	65.8 &	40.32\\
      & Margin & 55.57 & 47.01 & 47.52 & 56.28 &	68.34 &	54.03\\
      & Least Conf & 49.49 & 49.46 &	46.85 & 55.03 &	57.30 &	47.46 \\
      & DECAL Random & 61.44 &	54.5 &	48.48 & 64.16 &	63.42 &	46.37 \\
      & DECAL Entropy & 64.48 & 50.72 & 52.66 & 51.92 &	67.27 &	51.57\\
      & DECAL BADGE & 63.97 &	53.44 &	46.9 & 64.8 &	69.55 &	47.08 \\
      & DECAL Margin & 58.88 &	48.64 &	50.4 & 57.14 &	62.5 &	47.03 \\
      & DECAL Least Confidence & 60.48 &	49.30 &	49.78 & 60.06 &	64.42 &	40.99 \\
      \hline
      \midrule      
      \parbox[t]{2mm}{\multirow{10}{*}{\rotatebox[origin=c]{90}{DECAL Initialization}}} & Random & \cellcolor{yellow}\textbf{+9.21\%} &	\cellcolor{yellow}\textbf{+3.76\%} &	\cellcolor{yellow}\textbf{+5.95\%} & \cellcolor{yellow}\textbf{+2.03\%} &	-5.11\% &	-0.75\%\\
      & Entropy & -6.95\% &	\cellcolor{yellow}\textbf{+3.45\%} &	-10.74\% & \cellcolor{yellow}\textbf{+11.90\%} &	\cellcolor{yellow}\textbf{+24.73\%} &	\cellcolor{yellow}\textbf{+25.52\%} \\
      & BADGE & \cellcolor{yellow}\textbf{+5.73\%} & \cellcolor{yellow}\textbf{+8.64\%} &	\cellcolor{yellow}\textbf{+21.38\%} & \cellcolor{yellow}\textbf{+9.91\%} &	\cellcolor{yellow}\textbf{1.70\%} &	-2.48\% \\
      & Margin & -1.87\% &	\cellcolor{yellow}\textbf{+6.53\%} &	-16.45\% & -3.02\% &	-10.43\% &	\cellcolor{yellow}\textbf{+2.17\%} \\
      & Least Confidence & \cellcolor{yellow}\textbf{+17.61\%} &	-6.08\% &	-12.23\% & \cellcolor{yellow}\textbf{+5.23\%} &	\cellcolor{yellow}\textbf{+9.68\%} &	-6.27\% \\
        & DECAL Random & -1.17\% &	-6.01\% &	\cellcolor{yellow}\textbf{+14.35\%} & \cellcolor{yellow}\textbf{+4.09\%} &	\cellcolor{yellow}\textbf{+8.04\%} &	\cellcolor{yellow}\textbf{+22.88\%} \\      
      & DECAL Entropy & \cellcolor{yellow}\textbf{+4.67\%} &	-0.53\% &	-16.71\% & \cellcolor{yellow}\textbf{+4.31\%} &	-5.90\% &	\cellcolor{yellow}\textbf{+3.83\%} \\
      & DECAL BADGE & \cellcolor{yellow}\textbf{+3.25\%} &	-7.99\% &	\cellcolor{yellow}\textbf{+0.80\%} & \cellcolor{yellow}\textbf{+4.55\%} &	\cellcolor{yellow}\textbf{+1.01\%} &	\cellcolor{yellow}\textbf{+0.89\%} \\
      & DECAL Margin & \cellcolor{yellow}\textbf{+0.66\%} &	\cellcolor{yellow}\textbf{+5.53\%} &	\cellcolor{yellow}\textbf{+5.55\%} & \cellcolor{yellow}\textbf{+9.87\%} &	\cellcolor{yellow}\textbf{+2.91\%} &	\cellcolor{yellow}\textbf{+20.69\%} \\
      & DECAL Least Confidence & -4.19\% &	-2.86\% & -3.37\% & \cellcolor{yellow}\textbf{+3.15\%} &	-0.24\% &	\cellcolor{yellow}\textbf{+10.85\%} \\ 
      \bottomrule 
    \end{tabular}
    \end{adjustbox}
  \end{center}
\end{table}

\vspace{-3mm}
\vspace{-3mm}

\section{Conclusion and Next Steps}
In this work we motivate the design of a DECAL framework for medical applications. Augmenting active learning paradigms with EMR data creates the perfect setting for which active learning can be of true utility within the medical domain. In this study we have demonstrated this by designing an active learning framework constrained on a medically grounded prior gleaned from clinical EMR data about patient identity. 

Determining what other forms of EMR best serve active learning paradigms remains an open research question. Our next steps include investigating additional EMR data to inject into our framework for a more holistic analysis. Towards these efforts, we have collaborated alongside Retina Consultants of Texas (Houston, TX, USA) to create our own dataset (\cite{olives2022dataset}) that contains a plethora of EMR specific to ophthalmology. The clinical information consists of patient identity, general demographics, ocular disease state (Best Corrected Visual Acuity, Central Sub-field Thickness) and detailed ocular imaging in the form of spectral domain OCT, fundus photography and fluorescein angiography collected per the protocol. These were measured and recorded during routine visits to the clinic. In addition to DECAL, clinical context also aids research in supervised contrastive learning (\cite{kiran2022contrast1}).

\vspace{-3mm}
\acks{This material is based upon work supported by the National Science Foundation Graduate Research Fellowship under Grant No. DGE-1650044.}


\newpage




\vskip 0.2in
\bibliography{sample}

\end{document}